\documentstyle[12pt,psfig]{article}
\renewenvironment{thebibliography}[1]
 { \rm
   \begin{list}{\arabic{enumi}.}
    {\usecounter{enumi} \setlength{\parsep}{0pt}
     \setlength{\itemsep}{3pt} \settowidth{\labelwidth}{#1.}
     \sloppy
    }}{\end{list}}

\newcommand{\bo}{b_{1}(1235)}
\newcommand{\ft}{f_{2}(1270)}
\newcommand{\fo}{f_{1}(1285)}

\newcommand{\euq}{\eta_{u} (1295)}

\newcommand{\rqb}{\rho (1450)}

\def\k{{\bf k}}
\def\q{{\bf q}}
\def\p{{\bf p}}
\def\P{{\bf P}}

\def\lsim{\mathrel{\rlap{\lower4pt\hbox{\hskip1pt$\sim$}}
    \raise1pt\hbox{$<$}}}
\def\gsim{\mathrel{\rlap{\lower4pt\hbox{\hskip1pt$\sim$}}
    \raise1pt\hbox{$>$}}}

\begin{document}
\titlepage

\begin{flushright}
{IUNTC 01-02 \\}
{ADP-01-18/T453 \\}
{May 2001 \\}
\end{flushright}

\vglue 1cm

\begin{center}
{{\bf CHIRAL EXTRAPOLATIONS AND EXOTIC MESON SPECTRUM 
}
\vglue 1.0cm
{Anthony W. Thomas$^a$ and Adam P. Szczepaniak$^b$\\}

\bigskip
{\it $^a$ Special Research Centre for the Subatomic Structure of Matter
\\
and Department of Physics and Mathematical Physics, \\
Adelaide University, Adelaide SA 5005, Australia \\} 
\bigskip
{\it $^a$ Department of Physics and Nuclear Theory Center \\
 Indiana University, Bloomington, IN 47405, USA \\}}

\vglue 0.8cm

\end{center}

{\rightskip=2pc\leftskip=2pc\noindent
}

\vskip 1 cm
We examine the chiral corrections to exotic meson masses calculated in
lattice QCD. In particular, we ask whether the non-linear chiral
behavior at small quark masses, which has been found in other hadronic
systems, could lead to large corrections to the predictions of exotic meson 
masses based on linear 
extrapolations to the chiral limit. We find that our present
understanding of exotic meson decay dynamics suggests that  
open channels may not make a significant contribution to such non-linearities 
whereas the virtual, closed channels may be important. 
\newpage

\baselineskip=20pt

{\it 1. Introduction.} One of the biggest challenges in hadronic physics is to understand the role of 
gluonic degrees of freedom. Even though there is evidence from high energy 
experiments that gluons contribute significantly to hadron structure, for 
example to the momentum and spin sum rules, there is a pressing need 
{}for direct observation of gluonic excitations at low energies. 
In particular,  
it is expected that glue should manifest itself in the meson spectrum
and reflect on the nature of confinement.

In recent years a few candidates for glueballs and hybrid mesons have been
reported. For example, there are two major glueball candidates, the
$f_0(1500)$ and $f_0(1710)$, 
scalar-isoscalar mesons observed in ${\bar p}p$ annihilation, in central
production as well as in $J/\psi$ ($f_0(1710)$) decays~\cite{scalardata}. However, as 
a result of mixing with regular $q\bar q$ states,
 none of these is expected to
be a purely gluonic state. This is a general problem with 
glueballs -- they have
regular quantum numbers and therefore cannot be unambiguously identified
as purely gluonic excitations. On the other hand, the existence of mesons
with exotic quantum numbers (combinations of spin,
parity and charge 
conjugation, $J^{PC}$, which cannot be attributed to valence quarks alone), 
would be an explicit proof that gluonic degrees of freedom can indeed
play an active role at low energies.

Current estimates based on lattice QCD~\cite{latt,latt2}, 
as well as QCD-based models~\cite{mod}, 
suggest that the lightest exotic hybrid should have $J^{PC}=1^{-+}$ 
and mass slightly below 
$2 \mbox{ GeV}$.  On the experimental side, two exotic
candidates with these quantum numbers have recently been reported
by the E852 BNL collaboration.  One, in the $\eta\pi^-$ channel of the
reaction  $\pi^-(18\mbox{GeV})p \to X p \to \eta \pi^- p $, occurs at 
$M_X \approx 1400 \mbox{MeV}$~\cite{E8521}, while the other, in the 
$\rho^0\pi^-$ and $\eta'\pi^-$ channels,  
has a mass $M_X \approx 1600 \mbox{MeV}$~\cite{E8522,E8523}. Although the 
lighter state could be obscured by final state
interaction effects as well as leakage from 
the strong $X=a_2$ decay, the heavier one seems to have a fairly 
clean signal.

In the large $N_c$ limit it has been shown that exotic meson 
widths obey the same 
$N_c$ scaling laws as the regular mesons~\cite{cohen}.
The lack of an overwhelming number of 
exotic resonance candidates may therefore be the result of a small production 
cross-section in the typical reactions employed in exotic searches -- 
{\it e.g.,} 
high energy $\pi N$ or $K N$ scattering. Indeed, 
low lying exotic mesons are expected to have valence quarks in a spin-1 
configuration and their production would therefore 
be suppressed in peripheral pseudoscalar 
meson-nucleon scattering. By the same arguments one would expect exotic 
production to be enhanced in real photon-nucleon scattering~\cite{phth,phphen,AAAS}. 
This is particularly encouraging for the experimental studies of 
exotic meson photoproduction which have
recently been proposed in connection with 
the JLab $12\mbox{ GeV}$ energy upgrade~\cite{Alex}. 
 
In view of the theoretical importance of the topic and the
exciting new experimental possibilities for exotic meson searches it is of 
prime importance to constrain the theoretical predictions for exotic
meson masses -- especially from lattice QCD. One of the important
practical questions in this regard is the effect of light quarks on the
predicted spectrum. All calculations that have been performed so far
 involve quarks that are much heavier than the physical
$u$ and $d$ quarks and, with the exception of the initial study by the 
SESAM collaboration~\cite{latt2} (based on full QCD with 
two degenerate Wilson fermions), are performed in the quenched approximation. 
This is directly related to the present limitations on
computer power. It is well known that, as a consequence of dynamical
chiral symmetry breaking, hadron properties are non-analytic
functions of the light quark masses. This non-analyticity can lead to
rapid, non-linear variations of masses as the light quark masses
approach zero. Whether or not this could make a sizable difference 
to the lattice QCD predictions for exotic state masses is the question
we address.

There has been considerable activity concerning the appropriate
chiral extrapolation
of lattice results for hadron properties in the past few years,
ranging from magnetic moments \cite{MM} to charge radii \cite{CR} and 
structure functions \cite{SF} as well as masses \cite{MN}. The 
overall conclusion from this work is that for current quark masses
in the region above 50--60 MeV, where most lattice results are available, 
hadron properties are smoothly varying functions of the quark masses --
much like a constituent quark model. However, as one goes below 
this range, so that the corresponding pion Compton wavelength is larger
than the source, one finds rapid, non-linear variations which are a direct 
result of dynamical chiral symmetry breaking. These variations can 
change the mass of hadrons extracted by naive linear extrapolation by
a hundred MeV or more. (Certainly this was the case for the $N$ and 
$\Delta$ \cite{MN},
while for the $\rho$ the difference was only a few MeV \cite{RHO,LC}.) 
A variation of this order of magnitude for the predictions of exotic
meson masses would clearly be phenomenologically important and our
purpose is to check whether this seems likely. 

{\it 2. Chiral corrections from pion loops.} 
Our investigation of the possible non-linearity of the extrapolation of
exotic meson masses to the chiral limit, as a function of quark mass,
will follow closely  the earlier investigations for the $N$ and $\Delta$
baryons and the $\rho$ meson. The essential point is that rapid
non-linear variations can only arise from coupling to those pion-hadron
channels which give rise to  the leading (LNA) or next-to-leading 
non-analytic (NLNA) behavior of the exotic particle's self-energy. 
In a decay such as $E \rightarrow \pi H$,
this means that $H$ must be degenerate or nearly degenerate with the
exotic meson $E$. In addition, the relevant coupling constant should be
reasonably large. For example, in the case of the $\rho$ meson 
the relevant channels are $\omega \pi$ (LNA)
and $\pi \pi$ (NLNA) \cite{RHO} and the extrapolation of the $\rho$ meson mass
was carried out using:
\begin{equation}
m_\rho = a + b m_\pi^2 + \sigma_{\omega \pi} + \sigma_{\pi \pi},
\label{rho_extrap}
\end{equation}
where $\sigma_{i j}$ are the self-energy contributions from the channels
$i j = \omega \pi$ or $\pi \pi$.

Our first step in studying the extrapolation of exotic meson masses is
therefore to look at the channels to which the exotics can couple which
involve a pion and to check what is known about the corresponding
coupling constants. 
The matrix element describing a decay of the
$J^{PC}=1^{-+}$, isovector, exotic mesons with mass $m_X$, is given by: 
\begin{eqnarray}
& & \langle 1^{-+}; \P' |V| AB, \k, \P \rangle = 
(2\pi)^3\delta^3(\P'-\P) m_X \sum_{LS}\sum_{M_LM_S}Y_{LM_L}(\hat\k_X) 
\left( {k(m_X) \over m_X} \right)^L \nonumber \\
& & \times g_{LS}(k(m_X))  
 \langle s_A \lambda_A,s_B\lambda_B|SM_S\rangle 
\langle SM_S, LM_L|1M_X\rangle 
\langle I_A I^3_A, I_B I^3_B|1,I^3_X \rangle, 
\label{v}
\end{eqnarray}
where $k(m_X)=|\k|$ is the break-up momentum of the $AB$ meson pair, from a decay of a state of 
mass $m_X$, 
\begin{equation}
k(m_X) = \left( {{ (m_X^2 - (m_A - m_B)^2 )(m_X^2 - (m_A + m_B)^2 ) } \over 
{4m^2_X} } \right)^{1\over 2} \equiv \lambda(m_X,m_A,m_B), 
\end{equation}
produced with angular momentum $L$ and spin $S$, with 
${\bf L} + {\bf S} = {\bf 1}$.  

In terms of the couplings, $g_{LS}$, the partial decay widths are given by 
\begin{equation}
\Gamma_{LS}  = {m_X\over {32\pi^2}} \left( {k\over m_X} \right)^{2L+1} 
g^2_{LS}(k).
\end{equation}
These couplings have been calculated in Refs.~\cite{IKP,PSS}, in 
two models based on the flux tube picture of gluonic excitations but assuming 
different $q\bar q$ production mechanisms. 
In general the models agree on predicting sizable couplings to the
so called 
``S+P'' channels, where one of the two mesons in the final state has quarks in 
relative S-wave and the other in relative P-wave, {\it e.g.,} $\pi b_1$. 
In Table~1 we summarize the results of these calculations, listing only 
the decay channels containing a pion ($S$-wave quarks). 
Since the overall normalization of the 
decay matrix elements in these models is somewhat arbitrary, we have rescaled 
the original predictions given in Ref.~\cite{PSS} 
to match the total width 
($~170\mbox{ MeV}$) and mass ($m_X = 1.6\mbox{ GeV}$)  
of the exotic $\rho\pi$ state found by the E852 experiment~\cite{E8522}. 
\begin{table}
\begin{tabular}{|c|c|cc|cc|}
\hline
   Decay channel & wave & \multicolumn{2}{|c}{PSS}       & \multicolumn{2}{|c|}{IKP}      \\
                 &      &$g^2_{L}(k(m_X))/4\pi$&  $\Gamma_{L} [\mbox{ MeV}]$ & $g^2_{L}(k(m_X))/4\pi$&   $\Gamma_{L} [\mbox{ MeV}]$ \\ 
\hline
 $ \eta \pi    $ & P &  $9.4\times10^{-3}$ &    ${\cal O}(10^{-2})$ & $7.3\times10^{-3}$ & ${\cal O}(10^{-2})$     \\
 $ \eta^{'} \pi$ & P &  $9.4\times10^{-3}$ &    ${\cal O}(10^{-2})$ & $7.3\times10^{-3}$ & ${\cal O}(10^{-2})$     \\
$ \rho \pi    $ & P &   8.32              &    ${\cal O}(10)$ &  5.95
                 & ${\cal O}(10)$
\\
$ \ft\pi  $     & D &   5.3               &    ${\cal O}(10^{-2})$ &  \O                & \O  \\
$ \fo\pi  $     & S &   1.6               &    ${\cal O}(10)$ &  1.9
                 &  ${\cal O}(10)$    \\
                 & D &   2.3               &    ${\cal O}(10^{-2})$ &   309.             & ${\cal O}(1)$  \\
$ \bo\pi      $ & S &   6.1               &    ${\cal O}(100)$ &  6.54              &  ${\cal O}(100)$\\
                 & D &   37.9              &    ${\cal O}(1)$ &  324.
                 &   ${\cal O}(10)$  \\
$ \euq \pi $    & P &   37.6              &    ${\cal O}(10)$ & 21.25
                &  ${\cal O}(10)$ \\
$ \rqb \pi $    & P &   30.7              &    ${\cal O}(10^{-2})$ & 15.8               &  ${\cal O}(10^{-2})$ \\
\hline
\end{tabular}
\caption{Isovector Hybrid decay parameters }
\end{table}

Two features deserve particular attention. The couplings to ground state meson 
multiplets are generally smaller than to excited meson multiplets. This may be 
artificial; it is associated with the underlying structure 
of the model for pair creation which,  
when applied to exotic decays, strongly suppresses  
{}final states with mesons which have similar orbital wave functions. 
These models are based on a simple quark model picture 
in which (say) the $\pi$, $\rho$ 
and $\eta$ have very similar orbital wave functions. This is obviously an 
oversimplification.   
Secondly, the PSS and IKP models are drastically 
different when it comes to predicting 
ratios of branching ratios. In general, PSS predicts that 
higher partial waves should be strongly suppressed. 

An alternative approach was presented in Ref.~\cite{AAAS}. 
There, it was assumed 
that the decay of the observed exotic is dominated by just 
two channels : $\rho\pi$ -- the one in which it has been seen -- and the rest, 
say $b_1\pi$ (in an S-wave). The couplings obtained this way are 
shown in Table II.
\begin{table}
\begin{tabular}{|c|c|c|c|}
\hline
  Decay channel            &  wave  &     $g^2_L(k(m_X))/4\pi$ & $\Gamma_L [\mbox{
     MeV}]$      \\
\hline
 $ \rho \pi    $ & P &   24.59   & $O(100)$  \\
 $ \bo\pi      $ & S &   7.12    & $O(100)$    \\
\hline
\end{tabular}
\caption{Isovector Hybrid couplings $g^2_L(k(m_X))/4\pi$ }
\end{table}

The shift in the mass of the exotic meson associated with the 
pion loop, $\Sigma$, can
be calculated in second order perturbation theory (from $\Sigma  
= \langle \P'|VGV| \P'\rangle/ 
\langle \P'|\P \rangle$) and is given by:
\begin{eqnarray} 
\Sigma & = & \sum_{L} {{m_X^2} \over {64\pi^3}} {\cal P} \int_{m_A+m_B} {dM\over M}
\left( {{k(M)}\over m_X} \right)^{2L+1}   { {g^2_{L}(k(M)) }\over {m_X - M} }
\nonumber \\
& = &\sum_{L} {{m_X} \over {64\pi^3}} {\cal P} \int_0 {{dk k^2}
\over {\sqrt{k^2 + m_A^2}\sqrt{k^2 + m_B^2}}  }  
\left( {k\over m_X} \right)^{2L}   { {g^2_{L}(k) }\over {m_X - 
\sqrt{k^2 + m_A^2} -\sqrt{k^2 + m_B^2} } } \nonumber \\
\label{sigma}
\end{eqnarray}
with $k(M) = \lambda(M,m_A,m_B)$. 
The leading non-analytic (LNA) behavior of this self-energy is obtained by 
extracting the piece, independent of the 
ultra-violet cut-off (or form factor,  
$g_{L}(k)/g_{L}(k(m_X))$), which  
exhibits the strongest non-analytic 
variation as a function of the quark mass as one approaches the chiral
limit. (This is the term with the lowest odd power of $m_\pi$ or the
lowest power of $m_\pi$ multiplying $\ln m_\pi$.)
Setting $m_B = m_\pi$ the form of the LNA behavior 
depends on the relation between $m_X - m_A$ and $m_\pi$ and can be easily 
extracted analytically from Eq.~(\ref{sigma}) in two limiting cases. 
Consider first
the limit $m_X - m_A << m_\pi$, corresponding to an off-shell transition 
between the exotic meson and a nearly degenerate meson plus pion.  
In this case the 
self energy contribution reduces to 
\begin{equation}
\Sigma = -
\sum_{L} {1 \over {64\pi^3}} {\cal P} \int_0 {dk k^2}
\left( {k\over m_X} \right)^{2L}   { {g^2_{L}(k) }\over { 
 {k^2 + m_\pi^2} } }, \label{snr1}
\end{equation}
and the LNA behavior is given by 
\begin{equation}
\Sigma_{LNA} = \sum_L (-1)^L{{m_X}\over {32\pi}} {{g^2_{L}(k_\pi)} \over {4\pi}}  \left(
{{m_\pi} \over {m_X}} \right)^{2L+1}.
\label{Nr1}
\end{equation}
and $k_\pi \approx 0$.   
In the case $m_X - m_A >> m_\pi$, corresponding to a physical decay process 
to two light mesons (one of them being the pion), one obtains, 
\begin{equation}
\Sigma = 
\sum_{L} {{m_X} \over {64\pi^3}} {\cal P} \int_0 {dk k^2}
\left( {k\over m_X} \right)^{2L}   { {g^2_{L}(k) }\over { 
{m_A(m_X - m_A)\sqrt{k^2 + m_\pi^2} } } },
\label{snr2}
\end{equation}
leading to 
\begin{equation}
\Sigma_{LNA} = \sum_L (-1)^L{{m^3_X}\over {16\pi^2 m_A (m_X-m_A)}}
 {{(2L+1)!!}\over {(2L+2)!!}}{{g^2_{L}(k_\pi)} \over {4\pi}}  \left(
{{m^2_\pi} \over {m^2_X}} \right)^{L+1} \ln m_\pi.
\label{Nr2}
\end{equation}
Even though, in the present case, neither of these is the exact,
four-dimensional, pion loop contribution, they should 
give a reliable guide as to
the non-linearity that one may expect in the extrapolation of lattice
results to small quark mass.

In order to estimate the full self energy contribution, 
we need to know the off-shell dependence of the 
coupling constant. 
In the models of  Ref.~\cite{AAAS,IKP,PSS} the momentum dependence of the 
couplings arises from the Fourier transform of 
quark wave function overlaps and is typically of a 
gaussian form, 
\begin{equation}
g_{L}(k) = g_{L}(k(m_X))e^{-k^2/2\beta^2 + k^2(m_X)/2\beta^2} \label{ff}
\end{equation}
with the scale parameter expected to be in the range $\beta \approx 0.2-1\mbox{
GeV}$. To be consistent with 
the approximation of a heavy source, corresponding to 
Eqs.~(\ref{snr1}) and (\ref{snr2}), we need to set 
$k^2(m_X) \approx 0$  or $k^2(m_X) \approx 
(m_X - m_A)^2$ respectively. Thus 
the momentum dependence of the form factor near $k^2 = k^2_\pi$  
is not expected to significantly renormalize the value 
at $k^2 = k^2(m_X)$ 
and in Eqs.~(\ref{snr1}) and (\ref{snr2}) 
we can simply use the on-shell couplings  
{}from Tables 1 and 2. However, the off-shell 
transition matrix element may have a more significant momentum dependence. 
This could happen, for example, because of nodes present in the radial wave 
{}functions of excited mesons ({\it e.g.} the $\rho(1450)$ ). 
Such nontrivial momentum dependence of the form factor could possibly 
introduce a factor of $2-3$ uncertainty in our predictions. 
 Keeping this possibility in mind,  
using Eq.~(\ref{Nr1}) we find that for the maximal strength 
$P$-wave decays with couplings of the order of $20$,  
\begin{equation}
\Sigma \approx - 0.1 m_\pi^3 , \label{solp}
\end{equation} 
with both $\Sigma$ and $m_\pi$ in GeV, 
and all couplings of order 1 are irrelevant. 
For the $D$-waves with 
coupling as large as $300$ (IKP model, D-wave in the $b_1\pi$ 
channel) we obtain, 
\begin{equation}
\Sigma \approx + 0.5 m_\pi^5 ,\label{sold}
\end{equation} 
from Eq.(~\ref{Nr1}) or $\Sigma \approx + 0.4 m_\pi^6 \ln m_\pi$ from Eq.~(\ref{Nr2}). 
In the case of the $b_1\pi$ or $f_1\pi$ decay 
channels the second approximation is 
probably more accurate. 
\begin{figure}
\centerline{\psfig{figure=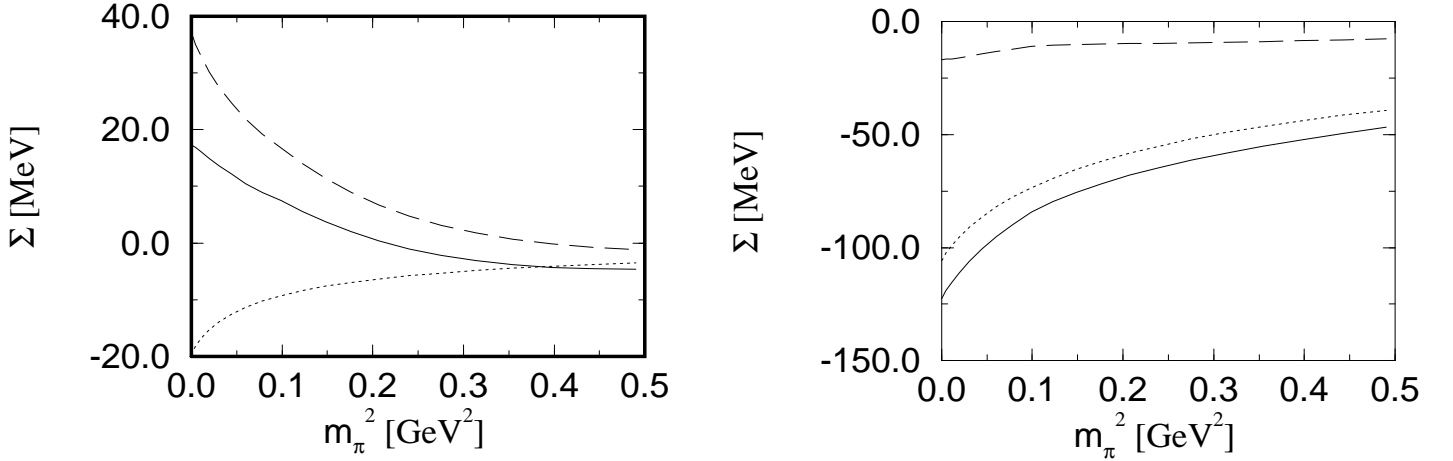,width=1.1\hsize}}
\smallskip
\caption{ Pion loop contribution to the exotic meson
 self energy from $P$-wave couplings calculated for 
 (a) $\beta=400\mbox{
 MeV}$ (left) and 
 (b) $\beta=700\mbox{ MeV}$ (right). The contributions
 from open and closed channels are shown with the dashed and dotted
 lines, {\it respectively} and the net correction is shown with the
 solid line. }
\label{fig1}
\end{figure}

\begin{figure}
\centerline{\psfig{figure=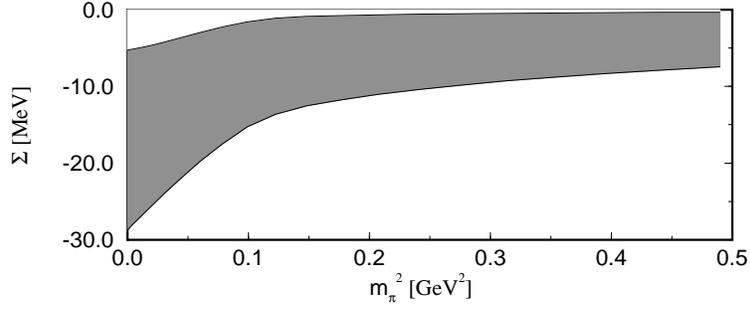,width=.7\hsize}}
\smallskip
\caption{ Pion loop contribution from the open, $D$-wave channels. The
  shaded region corresponds to $\beta$ in the range from $400$ to
  $700\mbox{ MeV}$ }
\label{fig2}
\end{figure}
To be more precise we also check how the full formula for the energy 
shift compares with the simpler expression for a heavy source given
above.
In Fig.~1 the dashed line shows $\Sigma$ as a function of the pion mass, 
{} calculated including the three largest $P$ wave open 
channels from Table~1 ($\rho(1450)\pi$, $\eta_u(1295)\pi$ and
$\rho\pi$) using the PSS
(larger) couplings.  The magnitude of $\Sigma$ is determined by the
 scale parameter chosen as $\beta=400\mbox{ MeV}$ in Fig.~1(a) and 
 $\beta=700\mbox{ MeV}$ in Fig~1(b). For small $\beta$ the dominant 
 contribution comes from the lightest open channel. This is  because 
 there is a large mismatch  between $k(m_X)$ and 
$\langle k \rangle \sim \beta$ which leads to enhancement from $g_L(k)$
 (Eq.~(\ref{ff})). As $\beta$ increases it is the channels with
 largest on-shell coupling that dominate. In Fig.~2 the contribution
 to $\Sigma$ from the two largest $D-wave$ couplings ($f_1\pi$ and
 $b_1\pi$ in the IKP model) is shown for $\beta$ in 
 the range between $400 \mbox{ MeV}$ and $700\mbox{ GeV}$.
In all cases we find that the mass shifts from the open channels do
  not exceed $50\mbox{
MeV}$. The magnitude of the self energy is naturally sensitive to the
scale parameter $\beta$, in particular for higher partial waves.
Nevertheless, the weak $m_\pi$ dependence for all $\beta$, in particular 
for the two limiting values corresponding to the upper and lower bounds, 
 decreasing for higher partial waves, 
is consistent with Eqs.~(\ref{solp}) and (\ref{sold}).
 
These are small shifts as compared, for example, to
the case of the pion contribution to the nucleon mass, where the LNA
term is of order $-5.6 m_\pi^3$. To be more specific, 
the $\pi N N$ vertex, written in the notation of Eq.~(\ref{v}), becomes 
\begin{equation}
g_{\pi NN} {\bar u}(\p',\lambda,I'_3)\gamma_5 \tau^{I^\pi_3} u(\p,\lambda,I_3) 
= m_N {\tilde g}_{\pi NN}
\langle {1\over 2}I_3,1I^\pi_3|{1\over 2} I'_3\rangle
\langle {1\over 2}\lambda,1M_L|{1\over 2}\lambda'\rangle  
{{q_\pi}\over  m_N}  Y_{1M_L}(\hat\q_\pi) 
\end{equation}
where 
\begin{equation}
{\tilde g}_{\pi NN} = \left( \sqrt{3}\times \sqrt{3} {{\sqrt{4\pi}}\over {\sqrt{3}}} 
g_{\pi NN}\right) 
\end{equation}
Eq.~(\ref{Nr1}) then gives, 
\begin{equation}
-{{m_N}\over {32\pi}}
{ { 3\times 4\pi g^2_{\pi NN}} \over {4\pi}} 
\left(  {{m_\pi} \over {m_X}} \right)^{3} 
= -{3\over {32\pi}} {{ g^2_{\pi NN}} \over {m_N^2}} 
m_\pi^3  = -{3\over {32\pi}} {{ g^2_A} \over {f_\pi^2}} 
m_\pi^3  \label{pnn}
\end{equation}
Comparing ${\tilde g}_{\pi NN}^2/4\pi = 12\pi \times 14.4 \approx 540$ to
the typical $P$-wave coupling from Table~1 or 2, it becomes clear why 
the open channels give small corrections to the exotic 
meson mass. 
 
However, just as in the case of the nucleon, one expects  
the largest LNA  behavior to come 
{}from virtual transitions to nearly degenerate states; in this case 
{}from transitions between an isovector exotic and 
a pion plus an isoscalar exotic meson. The couplings in Tables~1 and
2 only account for real decays and therefore cannot be used to 
estimate such transitions. To the best of our knowledge 
there are no microscopic calculations of  
such matrix elements, however, these can in principle be derived 
{}from PCAC. In the soft pion limit one has, 
\begin{eqnarray}
&& \langle 1^{-+}, \lambda', \p', I=1, a | q_\mu A^{\mu,b}(0) | 1^{-+},
\lambda, 
\p, I=0 \rangle = 
\nonumber \\
&& f_\pi \langle 1^{-+}, \p', \lambda',  I=1, a |V| \pi^b, \q ;
1^{-+}, \lambda, \p, I=0\rangle 
/ (2\pi)^3\delta^3(\p'-\p).
\end{eqnarray}
If one assumes the flux tube does not affect the 
axial charge, then in 
the static limit the valence quark contribution to the {\it lhs} is given by 
$2m_X\delta_{ab}{\vec\epsilon}^*(\lambda')\cdot[{\vec\epsilon}(\lambda)
\times {\vec q}]$. After comparing with Eq.~(\ref{v}) this yields 
\begin{equation}
g_{1^{-+}(I=1) \to 1^{-+}(I=0),\pi} = 2 \sqrt{ {8\pi}\over 3} {m_X\over
  f_\pi}. \label{gg} 
\end{equation}
The lack of any
contribution from the flux-tube makes this coupling identical to the
one for ordinary mesons, {\it e.g.} the  $\rho\omega\pi$ coupling (with
$m_X= m_\rho\sim m_\omega$). The phenomenological value for the
$g_{\rho\omega\pi}$, in our notation, is $g_{\rho\omega\pi}/m_\rho
= \sqrt{8\pi/3} {\tilde g}_{\rho\omega\pi}$ with 
${\tilde g}_{\rho\omega\pi} \sim 15\mbox{ GeV}^{-1}$, so that 
the simple quark model overestimates the coupling by approximately 
$50\%$. The flux-tube may contribute to the axial current if it 
couples to the spin of the quarks. In particular, since it is 
expected that the ground
state corresponds to a flux-tube in a $P$-wave with respect to the 
valence $Q{\bar Q}$ pair, the overlap of the spin and orbital
wave functions may 
modify the numerical factor in Eq.~(\ref{gg}). However, it is not 
expected to alter the $\sqrt{8\pi/3} m_X/f_\pi$ enhancement arising 
{}from the soft pion  emission. Thus, from Eqs.~(\ref{Nr1}) and ~(\ref{gg})
we estimate 
\begin{equation}
\Sigma = -{1\over {12\pi f_\pi^2}} m_\pi^3 \approx - 3 m_\pi^3. \label{nr}
\end{equation}

The magnitude of this correction is almost as large as 
that found in the nucleon case. 
The contribution to $\Sigma$ from the virtual transitions calculated
using the expression in Eq.~(\ref{snr1}), is shown in Fig.~1 with the
dotted line. Its magnitude is strongly dependent on the cutoff parameter
$\beta$, and
can be as large as ${\cal O}(100\mbox{MeV})$ for 
$\beta =700\mbox{ MeV}$. It also has significant variation as a function
of $m_\pi$, 
as expected from Eq.~(\ref{nr}).
The total self energy shift arising from the
real and  virtual corrections is shown in Fig.~1 with the solid line.
This simple analysis of virtual corrections also applies to the 
$\rho(1600)\pi$ channel which, depending on the relative phase, 
could enhance the overall LNA behavior by a factor of two. Finally if
the other light exotics $J^{PC}=0^{+-}, 2^{+-}$ are not too far from
the $J^{PC}=1^{-+}$, they could significantly enhance the 
$d$-wave ($\propto m_\pi^5$) behavior ({\it cf}. Fig.2).

{\it 3. Conclusions.}
We have explored the self-energy corrections to the mass of the $1^{-+}$
exotic meson which are most likely to introduce some 
non-linearity in the chiral extrapolation of lattice estimates of its mass.
As in earlier work involving the $N$ and $\Delta$ baryons and the $\rho$
meson, the guiding principle was to find those coupled channels 
involving a pion which yield the leading and next-to-leading 
non-analytic behavior of the 
self-energy. Using several models for the coupling constants to these
decay channels available in the literature we find that the 
effects of the open
channels are rather too small to lead to any significant non-linearity.

On the other hand, the virtual corrections can potentially be as large as 
those found for the 
nucleon, $\Delta$ or $\rho$ meson. 
Thus current estimates of the masses of exotic mesons, 
based on linear chiral extrapolation of
lattice results obtained at relatively large input quark masses, may not be
sufficiently accurate
and nonlinear behavior similar to that found for the nucleon
and $\Delta$ seems likely. A careful chiral extrapolation, constrained
by new calculations at lower quark mass (which are needed urgently), may
well lead to a physical mass for the $1^{-+}$ of order 100 MeV or more
lower than that found by naive linear extrapolation.
To estimate the strength of the virtual transition 
we used the soft pion limit theorem  
and in the future  more precise calculations 
of the axial current matrix elements involving 
exotic mesons should be performed to test this approximation. 
{}Finally, it will be essential to revisit these conclusions as we learn
more about the open channel exotic decay modes.  
In particular, if the existing models turn out to
significantly underestimate these pion couplings, then one would also need to
revise their contribution to the lattice extrapolation procedure. 

{\it 4. Acknowledgements}
We would like to acknowledge helpful discussions with P.~Page and
D.~Leinweber. 
This work was supported by the Australian Research Council and
the US Department of Energy under contract  DE-FG02-87ER40365. APS
would also like to acknowledge the support of the CSSM.

\vglue 0.4cm

\end{document}